\documentclass[dvips]{article}
\usepackage{icrctc07}

\title{VERITAS Data Acquisition}
\shorttitle{VERITAS Data Acquisition}

\authors{E. Hays$^{1,2}$ for the VERITAS Collaboration$^{3}$ }
\shortauthors{Hays, E. and et al.}
\afiliations{
$^1$University of Chicago, Enrico Fermi Institute, Chicago, Illinois 60637, U.S.A.\\ 
$^2$Argonne National Laboratory, Argonne, Illinois 60439, U.S.A.\\
$^3$For full author list see G. Maier, "The Status of VERITAS", these proceedings
}
\email{ehays@hep.anl.gov}

\abstract{
VERITAS employs a multi-stage data acquisition chain that extends from the VME
readout of custom 500 MS/s flash ADC electronics to the construction of telescope
events and ultimately the compilation of information from each telescope into array
level data. These systems provide access to the programming of the channel level
triggers and the FADCs. They also ensure the proper synchronization of event
information across the array and provide the first level of data quality monitoring.
Additionally, the data acquisition includes features to handle the readout of special
trigger types and to monitor channel scaler rates.  In this paper we describe the
software and hardware components of the systems and the protocols used to communicate
between the VME, telescope, and array levels. We also discuss the performance of the
data acquisition for array operations.
}

\begin{document}
\maketitle
\section{Introduction}

\begin{figure*}
\begin{center}
\includegraphics [width=0.8\textwidth]{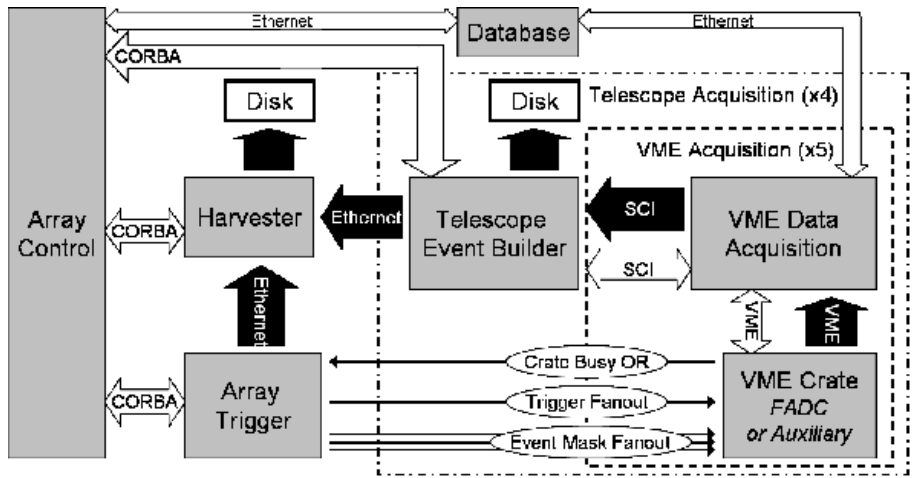}
\end{center}
\caption{Schematic of the data transfer and communication/control relationships for the data acquisition systems.  Thick black arrows indicate data transfer paths and the protocols used.  
Thick unfilled arrows indicate communication lines. 
The analog signals passed between the array trigger and the VME acquisition are included
 as thin arrows. 
The dashed line encloses processes repeated for each VME crate and the 
dash-dotted line encloses those repeated for each telescope.
}
\label{fig:daqdiag}
\end{figure*}

VERITAS \cite{icrc07:Veritas} is an array of 4 imaging Cherenkov telescopes designed to
record images of gamma rays impacting the atmosphere.
Photo-multiplier signals accompanying an image trigger \cite{icrc07:VeritasL3} 
are processed using a 3-tiered data acquisition system that 
operates at the VME crate, telescope, and array
levels.  The VME data acquisition guides the control and read out of
the electronics channels.  At the telescope level, the Event Builder
combines data from each VME crate into events.  At the array level,
the Harvester collects events from each telescope and the array
trigger and forms the final data product. 
Figure \ref{fig:daqdiag} shows a schematic of the data flow
and communication between the processes.

\subsection{The VME Data Acquisition}

The VME Data Acquisition (VDAQ) serves as the interface to 
five VME crates that participate in the
digitization of the PMT signals and the channel-level triggering for each telescope. 
Four of
these contain 500 MS/s flash ADC modules with a clock-trigger 
module \cite{icrc03:VeritasFADC} and a fifth serves as an
auxiliary crate housing a specialized clock-trigger module and a GPS
clock (TTM637VME).  The FADC electronics and the constant fraction
discriminators (CFDs)\cite{icrc03:VeritasCFD} that produce individual
channel triggers are housed on 10-channel 9U VME modules.  The
required complement of FADC channels to accomodate the 499-pixel
camera are distributed among 4 crates of 12 or 13 FADC modules. 

Each crate is controlled by a VMIVME 7807 Intel Pentium M 1.8 GHz single
board computer running Linux.  An additional Dolphin PCI
mezzanine card provides a communication link using the ANSI/IEEE
1596-1992 Scaleable Coherent Interface (SCI) standard.  Each of the
VME crates and the telescope Event Builder are connected
as nodes of a low-latency, high-throughput network. The configuration 
used in VERITAS achieves transfer rates of 50 MBytes/s.
Data transfers and general communication between VDAQ and the
telescope Event Builder are conducted via this interface. Each crate
can be accessed via GBit ethernet for starting and halting the
acquisition process.

A 24-sample (48 ns) FADC trace translates into an event fragment size
of 3880 bytes for a crate of 13 modules (130 channels). Event
fragments are collected and buffered until the accumulated size
reaches 8 MBytes ($\sim$2200 events for FADC crates).  The buffer is
then shipped to the telescope Event Builder.  The transfers from the
VME crates happen asynchronously for crates containing different
numbers of modules. The typical data rate for a crate is $\sim$780
kBytes/s, which is well below the maximum rates permitted by the SCI
and the Event Builder.  The chief limitation on the system throughput
arises from the data transfer rate over the VME backplane.

VDAQ is an event-driven process while the telescope and array
acquisition processes are buffered.  The telescope deadtime is
incurred only at the VME level and is dominated by the size of the
crate event fragments. For an array trigger rate of 200 Hz, the
deadtime at an individual telescope is about 8.5\%.

{\bf Hardware Configuration.}  Upon initiation, the VDAQ process for
each crate generates a physical map of the modules present by type
and, if applicable, a unique board identifier.  Each crate provides
this map to the Event Builder and uses it internally to procure
hardware configuration parameters for each FADC channel and CFD.  
A variety of programmable features of
the FADCs and CFDs are configured \cite{icrc03:VeritasFADC,icrc03:VeritasCFD}.
While some parameters are unique to an individual physical component
and must be associated by a uniqe identifier, others are properties of
the telescope and pixel to which a channel is connected.  The
configuration settings and component mappings are stored in a MySQL
database. This solution suits the distributed nature of the crate
acquisition and accomodates the mappings required to properly
configure the hardware.  An additional benefit is an accurate log of
the mappings and settings used.

{\bf Trigger Synchronization.} After initialization and configuration,
the acquisition processes await signals from the array trigger or
commands from the Event Builder.  The clock-trigger module in each
crate generates a busy level during FADC reads when triggers cannot be
accepted. To keep the five-crate system aligned, the busy levels are
combined into a single telescope busy level and applied locally as a
veto to incoming triggers.  This signal is sent to the array trigger
\cite{icrc07:VeritasL3} to indicate when the telescope cannot accept
triggers.

{\bf Trigger Type Handling.} Upon receipt of an array trigger, each
crate is passed a serialized event mask that contains an event number
and a trigger type code.  The mask is encoded into the event fragment
via the clock-trigger boards to allow sychronization by the Event
Builder. The trigger code is read by VDAQ and can be used to indicate
special requests for FADC functions from the array trigger.
Out-of-time reads of the FADC memory buffer are used to assess
pedestal values periodically during observations.  Upon receipt of a
pedestal trigger code from the array trigger, VDAQ invokes a dedicated
hardware command in the FADCs.  Pedestal events are included in the
data as normal events distinguished by their type code.

{\bf Trigger Rate Measurements.} VDAQ accesses CFD trigger rates 
via the FADC modules.  Scaler counts for each channel are
read every 400 events, about once every 2 seconds; this is not often
enough to impact the deadtime significantly.  The CFD scalers are
packed as a specially tagged event and shipped to the Event Builder as
part of the regular SCI transfer.  Scaler reads are included during
normal observations to provide direct diagnostics of channel-level
triggering.

\subsection{The Telescope Event Builder}
Each VERITAS telescope has a dedicated Telescope Event Builder which
is responsible for combining the event fragments from each of the five
VME crates to produce telescope events; these are then written to
local disk and sent via GBit ethernet to the Harvester system.
The Telescope Event-Building system is a Dual Intel Xeon
Server machine running linux and using a local RAID array. 
Communications with and control
of the Event Builder program is achieved through use of CORBA (Common
Object Request Broker Architecture, specifically OmniORB). The Event
Building Software is written in C++ and is fully multithreaded,
typically containing five threads whilst running: Communications, SCI
Buffer Acquisition and Parsing, Event Building, Disk Writer, Network
Writer.

At the start of each night, the Event Builder queries each VDAQ crate for a
map of present VME modules and then dynamically configures itself.
Data is buffered at each of the VDAQ machines and transferred in
blocks to the Event Builder via the SCI system. The Event Builder
parses these memory blocks and extracts pieces of individual events,
tagged by unique event numbers, which are then stored in memory. When
all of the pieces of an event have arrived, the telescope event is
built and buffered. Once roughly 160 kBytes of telescope events have
been accumulated, the events are then dispatched to the
``Consumers''. The Consumers are processes that receive built data
buffers. They utilize a common architecture and, at run-time, any
number of consumers may be registered. Typically only two are; the Disk
Writer and Network Writer, but an additional Data Integrity Monitor
consumer may also be used.  It is estimated that the throughput of the
Telescope Event-Building system on a dual 2.4 GHz Xeon server is
approximately 12 MBytes/s. In addition to receiving actual event data,
the Telescope Event Builders periodically receive CFD scaler
data as described in the previous section. These data are not
transferred with the event data, but simply stored and made available
via CORBA to any system that requests the information.

\subsection{The Harvester}

VERITAS back-end data acquisition is the task of the Harvester -- a single
eight-core machine that collects data from all telescopes in real-time
in addition to a stream of meta-data from the L3 trigger.  The Harvester
accomplishes four tasks:

{\bf Storing data.}  The Harvester saves all data streams to
its fast RAID in real time.  The current strategy for real-time data
storage is to create a separate file for each telescope; this makes it
possible to handle the separate telescope streams in parallel with minimal
interaction.

{\bf Combining data.}  In addition to saving the data, the
Harvester combines it in real time into \emph{array events}.  In this way,
real time diagnostics can see a ``big picture'' of what the array is doing,
rather than having to deal with telescopes individually.

{\bf Diagnostics.}  The Harvester runs a variety of real-time
diagnostics -- ranging from sanity checks to see if telescopes read out when
they were supposed to, to a complete high-performance stereo analysis package
that serves as the VERITAS real-time quicklook.  During a run,
the observer has up-to-the-second knowledge of the performance of the system.

{\bf Creating the final product.}  After a run completes, the
Harvester immediately starts combining the data streams from the run to create
a single file using the VERITAS Bank File (VBF) data format.

\smallskip
VBF groups telescope events together, such that given an event number,
the user has immediate access to the corresponding events from all
telescopes, in addition to meta-data from the array trigger.  Thus,
the analysis does not need to correlate stereo events; this task is
already accomplished by the data acquisition system.

VBF has been designed for portability, high performance reading and
writing, compactness, and extensibility.  High performance access and
compactness are achieved using a custom scheme for compressing FADC
samples based on picking a different number of bits-per-sample
depending on the dynamic range of each particular trace.  This scheme
overwhelmingly outperforms gzip and bzip2 in reading and writing times
while reaching similar compactness. See Table~\ref{vbfperf} for
performance measurements

\begin{table}
\begin{center}
\begin{tabular}{|l|c|c|}
\hline
VBF Variant & Space Usage & Read Time
\\
\hline
\hline
Uncompressed                & 100\% & 100\% \\ 
w/ Gzip                        & 42\%  & 114\% \\ 
w/ Bzip2                       & 35\%  & 514\% \\ 
\hline
Compressed$^*$ & 38\%  & 64\% \\ 
Comp. w/ Gzip  & 32\%  & 93\% \\ 
Comp. w/ Bzip2 & 30\%  & 471\% \\ 
\hline
\end{tabular}
\end{center}
\caption{Comparison of compression schemes normalized to the uncompressed VBF case.
performed on a Pentium 4 with 2 GB RAM, an eight-way SCSI RAID-0, and Linux 2.4.18.
For bzip2, we used a block size of 100,000 bytes.
$^*$VERITAS uses custom compression alone. 
}
\label{vbfperf}
\end{table}

The observer interacts with the Harvester using the VERITAS array
control software, as well as a suite of GUIs designed to view the
results of quicklook analysis. 
The stereo analysis performed by the VERITAS quicklook system is
capable of a very high level of performance -- both in time and in
sensitivity.  A typical twenty minute run takes two minutes to analyze
using quicklook.
Further, the quicklook analysis results are
comparable to offline analysis packages.  Thus, we have confidence
that if a bright source appeared in our field of view during
observations, quicklook would have as good of a chance of seeing it as
a manually performed offline analysis.

\section{Conclusion}

The VERITAS data acquisition systems combine a variety of hardware and
software resources to achieve efficient and reliable operation from
the reading of FADC data to the building and storage of event data.
Diagnostic information is available at all levels and real-time
analsyis is performed to ensure data quality. The system has proven
highly adaptable and meets the needs of a variety of configuration and
calibration tasks. All systems operate within design parameters and
leave room for the exploration of low-threshold regimes.

\subsection*{Acknowledgments}

This research is supported by grants from the U.S. Department of Energy,
the U.S. National Science Foundation,
and the Smithsonian Institution, by NSERC in Canada, by PPARC in the UK and
by Science Foundation Ireland.

\bibliography{icrc1166}
\bibliographystyle{plain}
\end{document}